%% file: RMT26.tex
\documentclass[11pt]{book}

\usepackage{amssymb}
\usepackage{amsmath,amsthm}
\usepackage{amsfonts}
\usepackage{epsfig}

\usepackage{graphicx}
\usepackage{color}
\usepackage[utf8]{inputenc}
\usepackage{xspace}
\usepackage{hyperref}

\makeatletter

\renewenvironment{thebibliography}[1]
     {\section*{\bibname}%
      \@mkboth{\MakeUppercase\bibname}{\MakeUppercase\bibname}%
      \list{\@biblabel{\@arabic\c@enumiv}}%
           {\settowidth\labelwidth{\@biblabel{#1}}%
            \leftmargin\labelwidth
            \advance\leftmargin\labelsep
            \@openbib@code
            \usecounter{enumiv}%
            \let\p@enumiv\@empty
            \renewcommand\theenumiv{\@arabic\c@enumiv}}%
      \sloppy
      \clubpenalty4000
      \@clubpenalty \clubpenalty
      \widowpenalty4000%
      \sfcode`\.\@m}
     {\def\@noitemerr
       {\@latex@warning{Empty `thebibliography' environment}}%
      \endlist}

\makeatother

\newcommand{\sect}[1]{\setcounter{equation}{0}\section{#1}}

\renewcommand\bibname{References}

\newcommand{\define}{\emph}
\newtheorem{theorem}{Theorem}
\newtheorem*{corollary}{Corollary}
\newtheorem*{remark}{Remark}

\DeclareMathOperator{\Tr}{Tr}
\newlength{\intwidth}

\DeclareMathAlphabet{\mathitbf}{OML}{cmm}{b}{it}

\newcommand{\CicutaMolinari}{14\xspace}

\newcommand{\Orantin}{16\xspace}
\newcommand{\ZinnZuber}{27\xspace}
\newcommand{\Kostov}{30\xspace}

\begin{document}


\setcounter{chapter}{25}

\chapter{Enumeration of maps}
\thispagestyle{empty}

\ \\

\noindent
{{\sc J. Bouttier}
\\~\\
Institut de Physique Th\'eorique, CEA Saclay,\newline 
F-91191 Gif-sur-Yvette Cedex, France}

\begin{center}
{\bf Abstract}

This chapter is devoted to the connection between random matrices and
maps, i.e graphs drawn on surfaces. We concentrate on the one-matrix
model and explain how it encodes and allows to solve a map
enumeration problem.
\end{center}

\sect{Introduction}
\label{sec:intro}

Maps are fundamental objects in combinatorics and graph theory,
originally introduced by Tutte in his series of ``census'' papers
\cite{Tut62a,Tut62b,Tut62c,Tut63}. Their connection to random matrix
theory was pioneered in the seminal paper titled ``Planar diagrams''
by Brézin, Itzykson, Parisi and Zuber \cite{BIPZ} building on an
original idea of 't Hooft \cite{tH74}. In parallel with the idea that
planar diagrams (i.e maps) form a natural discretization for the
random surfaces appearing in 2D quantum gravity (see chapter \Kostov),
this led to huge developments in physics among which some highlights
are the solution by Kazakov \cite{Ka86} of the Ising model on
dynamical random planar lattices (i.e maps) then used as a check for
the Knizhnik-Polyakov-Zamolodchikov relations \cite{KPZ}.

Matrix models, or more precisely \emph{matrix integrals}, are
efficient tools to address map enumeration problems, and complement
other techniques which have their roots in combinatorics, such as
Tutte's original recursive decomposition or the bijective
approach. Despite the fact that these techniques originate from
different communities, it is undoubtable that they are intimately
connected. For instance, it was recognized that the loop equations for
matrix models correspond to Tutte's equations \cite{Eyn06}, while the
knowledge of the matrix model result proved instrumental in finding a
bijective proof \cite{census} for a fundamental result of map
enumeration \cite{BeCa94}.

In this chapter, we present an introduction to the method of matrix
integrals for map enumeration. To keep a clear example in mind, we
concentrate on the problem considered in \cite{BeCa94,census} of
counting the number of planar maps with a prescribed degree
distribution, though we mention a number of possible generalizations
of the results encountered on the way. Of course many other problems
can be addressed, see for instance chapter \Kostov for a review of
models of statistical physics on discrete random surfaces (i.e maps)
which can be solved either exactly or asymptotically using matrix
integrals.

This chapter is organized as follows. In section \ref{sec:mapdefs}, we
introduce maps and related objects. In section \ref{sec:feynman}, we
discuss the connection itself between matrix integrals and maps: we
focus on the Hermitian one-matrix model with a polynomial potential,
and explain how the formal expansion of its free energy around a
Gaussian point (quadratic potential) can be represented by diagrams
identifiable with maps. In section \ref{sec:onemmsol}, we show how
techniques of random matrix theory introduced in previous chapters
allow to deduce the solution of the map enumeration problem under
consideration. We conclude in section \ref{sec:bij} by explaining how
to translate the matrix model result into a bijective proof.


\sect{Maps: definitions}
\label{sec:mapdefs}

Colloquially, maps are ``graphs drawn on a surface''. The purpose of
this section is to make this definition more precise. We recall the
basic definitions related to graphs in subsection \ref{sec:graphs},
before defining maps as graphs embedded into surfaces in subsection
\ref{sec:mapembed}. Subsection \ref{sec:combimaps} is devoted to a
coding of maps by pairs of permutations, which will be useful for the
following section.

\subsection{Graphs}
\label{sec:graphs}

A \define{graph} is defined by the data of its \define{vertices}, its
\define{edges}, and their \define{incidence relations}. Saying that a
vertex is incident to an edge simply means that it is one of its
extremities, hence each edge is incident to two vertices. Actually, we
allow for the two extremities of an edge to be the same vertex, in
which case the edge is called a loop. Moreover several edges may have
the same two extremities, in which case they are called multiple
edges. A graph without loops and multiple edges is said to be
\define{simple}. (In other terminologies, graphs are always simple and
multigraphs refer to the ones containing loops and multiple edges.) A
graph is \define{connected} if it is not possible to split its set of
edges into two or more non-empty subsets in such a way that no edge is
incident to vertices from different subsets. The \define{degree of a
  vertex} is the number of edges incident to it, counted with
multiplicity (i.e a loop is counted twice). 

\subsection{Maps as embedded graphs}
\label{sec:mapembed}

\begin{figure}[htbp]
  \begin{center}
    \includegraphics[width=\textwidth]{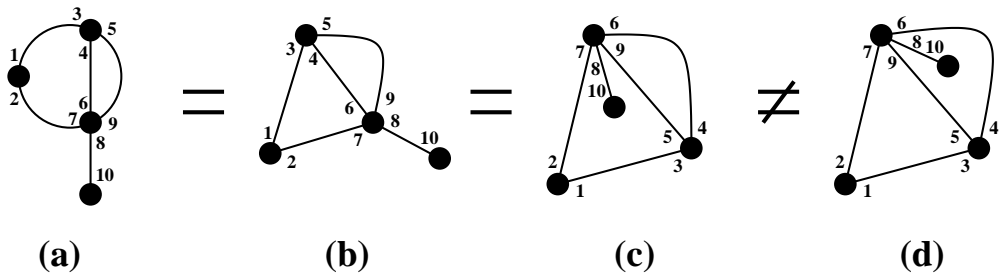}
    \caption{Several cellular embeddings of a same graph into the
      sphere (drawn in the plane by stereographic projection), where
      half-edges are labelled from 1 to 10. (a) and (b) differ only by
      a simple deformation. (c) is obtained by choosing another pole
      for the projection. Thus (a), (b) and (c) represent the same
      planar map encoded by the permutations
      $\sigma=(1\,2)(3\,4\,5)(6\,7\,8\,9)(10)$ and
      $\alpha=(1\,3)(2\,7)(4\,6)(5\,9)(8\,10)$. (d) represents a
      different map, encoded by the permutations $\sigma'=\sigma \circ
      (8\,9)$ and $\alpha'=\alpha$. Their inequivalence may be checked
      by comparing the face degrees or noting that $\sigma \circ
      \alpha$ is not conjugate to $\sigma' \circ \alpha'$.}
    \label{fig:planarmaps}
  \end{center}
\end{figure}

The surfaces we consider here are supposed to be compact, connected,
orientable\footnote{It is however possible to extend our definition to
  non-orientable surfaces.} and without boundary. It is well-known
that such surfaces are characterized, up to homeomorphism, by a unique
nonnegative integer called the genus, the sphere having genus 0.

An \define{embedding} of a graph into a surface is a function
associating each vertex with a point of the surface and each edge with
a simple open arc, in such a way that the images of distinct graph
elements (vertices and edges) are disjoint and incidence relations are
preserved (i.e the extremities of the arc associated with an edge are
the points associated with its incident vertices). The embedding is
\define{cellular} if the connected components of the complementary of
the embedded graph are simply connected. In that case, these are
called the \define{faces} and we extend the incidence relations by
saying that a face is incident to the vertices and edges on its
boundary. Each edge is incident to two faces (which might be equal)
and the \define{degree of a face} is the number of edges incident to
it (counted with multiplicity). It is not difficult to see that the
existence of a cellular embedding requires the graph to be connected
(for otherwise, graph components would be separated by
non-contractible loops).

\define{Maps} are cellular embeddings of graphs considered up to
continuous deformation, i.e up to an homeomorphism of the target
surface.  Clearly such a continuous deformation preserves the
incidence relations between vertices, edges and faces. It also
preserves the genus of the target surface, hence it makes sense to
speak about the genus of a map. A \define{planar map} is a map of
genus 0. In all rigor, this notion should not be confused with that of
planar graph\footnote{Unfortunately the literature is often not that
  careful. To illustrate the distinction let us mention that, while
  the first enumerative results for maps date back to Tutte in the
  1960's, the mere asymptotic counting of planar graphs (in our
  present definition) is a fairly recent result \cite{GiNo05}.}, which
is a graph having a least one embedding into the sphere. Figure
\ref{fig:planarmaps} shows a few cellular embeddings of a same
(planar) graph into the sphere (drawn in the plane by stereographic
projection): the first three are equivalent to each other, but not to
the fourth (which might immediately be seen by comparing the face
degrees). Let us conclude this section by stating the following
well-known result.

\begin{theorem}[Euler characteristic formula]
  In a map of genus $g$, we have
  \begin{equation}
    \#\{\mathrm{vertices}\} - \#\{\mathrm{edges}\} +
    \#\{\mathrm{faces}\} = 2 - 2 g.
  \end{equation}
  This quantity is the \define{Euler characteristic} $\chi$ of the map.
\end{theorem}

In the planar case $g=0$, the formula is also called the Euler
relation.

\subsection{Combinatorial maps}
\label{sec:combimaps}

As discussed above, a map contains more data than a graph, and one may
wonder what is the missing data. It turns out that the embedding into
an oriented surface amounts to defining a cyclic order on the edges
incident to a vertex\footnote{This observation is generally attributed
  to Edmonds \cite{Ed60}. For a comprehensive graph-theoretical
  treatment of embeddings into surfaces, we refer to the book of Mohar
  and Thomassen \cite{MoTh01}.}. More precisely, as we allow for
loops, it is better to consider \define{half-edges}, each of them
being incident to exactly one vertex. Let us label the half-edges by
distinct consecutive integers $1,\ldots,2m$ (where $m$ is the number
of edges) in an arbitrary manner. Given a half-edge $i$, let
$\alpha(i)$ be the other half of the same edge, and let $\sigma(i)$ be
the half-edge encountered after $i$ when turning counterclockwise
around its incident vertex (see again Figure \ref{fig:planarmaps} for
an example with $m=5$). It is easily seen that $\sigma$ and $\alpha$
are permutations of $\{1,\ldots,2m\}$, which furthermore satisfy the
following properties:
\begin{itemize}
\item[(A)] $\alpha$ is an involution without fixed point,
\item[(B)] the subgroup of the permutation group generated by $\sigma$
  and $\alpha$ acts transitively on $\{1,\ldots,2m\}$.
\end{itemize}
The latter property simply expresses that the underlying graph is
necessarily connected. The subgroup generated by $\alpha$ and $\sigma$
is called the cartographic group \cite{Zv95}.

These permutations fully characterize the map. Actually, a pair of
permutations ($\sigma,\alpha$) satisfying properties (A) and (B) is
called a \define{labelled combinatorial map} and there is a one-to-one
correspondence between maps (as defined in the previous section) with
labelled half-edges and labelled combinatorial maps. In this
correspondence, the vertices are naturally associated with the cycles
of $\sigma$, the edges with the cycles of $\alpha$ and the faces with
the cycles of $\sigma \circ \alpha$. At this stage one may have
noticed that vertices and faces play a symmetric role, indeed
$(\sigma \circ \alpha,\alpha)$ is also a labelled combinatorial map
corresponding to the \define{dual map}. Degrees are given the
length of the corresponding cycles, and the Euler characteristic is
given by
\begin{equation}
  \label{eq:eulerperm}
  \chi(\sigma,\alpha) = c(\sigma) - c(\alpha) + c(\sigma \circ \alpha)
\end{equation}
where $c(\cdot)$ denotes the number of cycles.

Our coding of maps via pairs of permutations depends on an arbitrary
labelling of the half-edges, which might seem unsatisfactory. We shall
identify configurations differing by a relabelling, and it easily seen
that this amounts to identifying $(\sigma,\alpha)$ to all $(\rho \circ
\sigma \circ \rho^{-1},\rho \circ \alpha \circ \rho^{-1})$ where
$\rho$ is an arbitrary permutation of $\{1,\ldots,2m\}$. These
equivalence classes are in one-to-one correspondence with unlabeled
maps. This distinction has some consequences for enumeration: for
instance the number of labelled maps with $m$ edges is \emph{not}
equal to $(2m)!$ times the number of unlabeled maps, because some
equivalence classes have fewer than $(2m)!$ elements (due to
symmetries). By the orbit-stabilizer theorem, the number of elements
in the class of $(\sigma,\alpha)$ is $(2m)!/\Gamma(\sigma,\alpha)$
where $\Gamma(\sigma,\alpha)$ is the number of permutations $\rho$
such that $\sigma=\rho \circ \sigma \circ \rho^{-1}$ and $\alpha=\rho
\circ \alpha \circ \rho^{-1}$ (such $\rho$ is an
\define{automorphism}). $\Gamma(\sigma,\alpha)$ is often called the
``symmetry factor'' in the literature, though it is seldom properly
defined.

As we shall see in the next section, matrix integrals are naturally
related to labelled maps. Enumerating unlabeled maps is a harder
problem as it requires classifying their possible symmetries, which is
beyond the scope of this text\footnote{For more on the topic of
  combinatorial maps and their automorphisms, we refer the reader to
  \cite{Co75,CoMa92} and references therein. \cite{CoMa92} also
  discusses hypermaps, which are the natural generalization of
  combinatorial maps obtained when relaxing constraint (A): these are
  actually bipartite maps in disguise, associated with the two-matrix
  model of equation (\ref{eq:twomatfree}).}. The distinction is
circumvented when considering \define{rooted} maps i.e maps with a
distinguished half-edge (often represented as a marked oriented edge):
such maps have no non-trivial automorphism hence the enumeration
problem is equivalent in the labelled and unlabeled case. Most
enumeration results in the literature deal with rooted
maps. Furthermore maps of large size are ``almost surely''
asymmetric, so the distinction is irrelevant in this context.

\sect{From matrix integrals to maps}
\label{sec:feynman}

In this section, we return to random matrices in order to present
their connection with maps. We will concentrate on the so-called
one-matrix model (though we will allude to its generalizations): maps
appear as diagrams representing the expansion of its partition
function around a Gaussian point. Our goal is to explain this
construction in some detail, as this might be also useful for the
comprehension of other chapters.

This section is organized as follows. Subsection \ref{sec:onemmdef}
provides the definitions and the main statement (Theorem
\ref{thm:freetopo}) of the topological expansion in the one-matrix
model. The following subsections are devoted to its derivation and
generalizations. Subsection \ref{sec:gaussian} discusses Wick's
theorem for Gaussian matrix models. Subsection \ref{sec:diagrams}
introduces \emph{ab initio} the diagrammatic expansion of the
one-matrix model. Subsection \ref{sec:matrixcombi} formalizes this
computation and shows its natural relation with combinatorial
maps. Subsection \ref{sec:multimat} finally presents a few
generalizations of the one-matrix model.

\subsection{The one-matrix model}
\label{sec:onemmdef}

We consider the model of a Hermitian random matrix in a polynomial
potential, often called simply the \define{one-matrix model}, which
has been already discussed in previous chapters. Different notations
and conventions exist, for the purposes of this chapter we define its
partition function as
\begin{equation}
  \label{eq:perturb}
  \Xi_N(t,V) = \frac{\int e^{N \Tr \left(- M^2/(2t) +
        V(M)\right)} \, dM}{\int e^{-N \Tr M^2/(2t)}\, dM}
\end{equation}
where $dM$ is the Lebesgue measure over the space of
Hermitian matrices, and $V$ stands for a ``perturbation'' of
the form
\begin{equation}
  \label{eq:potential}
  V(x) \equiv \sum_{n=1}^{\infty} \frac{v_n}{n} x^n.
\end{equation}
Here we shall consider the coefficients $v_n$ as formal variables
hence, rather than a polynomial, $V$ is a formal power series in $x$
and the $v_n$. In this sense, a more \define{proper definition of the
  partition function} $\Xi_N(t,V)$ is
\begin{equation}
  \label{eq:onemmpfdef}
  \Xi_N(t,V) = \left\langle e^{N \Tr V(M)} \right\rangle
\end{equation}
where $\langle \cdot \rangle$ denotes the expectation value with
respect to the Gaussian measure proportional to $e^{-N \Tr
  M^2/(2t)}\, dM$, acting coefficient-wise on $e^{N \Tr V(M)}$ viewed as a
formal power series in the $v_n$ whose coefficients are polynomials in
the matrix elements. In other words, the matrix integral in
(\ref{eq:perturb}) must be understood in the ``formal'' sense of
chapter \Orantin. We define furthermore the \define{free energy} by
\begin{equation}
  \label{eq:freeenerdef}
  F_N(t,V) = \log \Xi_N(t,V).
\end{equation}
The main purpose of this section is to establish the following
theorem, which is essentially a formalization of ideas present in
\cite{BIPZ,tH74}.
\begin{theorem}[Topological expansion]
  \label{thm:freetopo}
  The free energy of the one-matrix model has the ``topological''
  expansion
  \begin{equation}
    \label{eq:topological}
    F_N(t,V) = \sum_{g=0}^{\infty} N^{2-2g} F^{(g)}(t,V)
  \end{equation}
  where $F^{(g)}(t,V)$ is equal to the exponential generating function
  for labelled maps of genus $g$ with a weight $t$ per edge and, for
  all $n\geq 1$, a weight $v_n$ per vertex of degree $n$.
\end{theorem}

\begin{corollary}
  The quantity
  \begin{equation}
    \label{eq:rootedgf}
    E^{(g)}(t,V) = 2 t \frac{\partial F^{(g)}(t,V)}{\partial t}
  \end{equation}
  is the generating function for rooted maps (i.e maps with a
  distinguished half-edge) of genus $g$. Similarly $n v_n
  \frac{\partial F^{(g)}(t,V)}{\partial v_n}$ corresponds to rooted
  maps of genus $g$ whose root vertex (i.e the vertex incident to the
  distinguished half-edge) has degree $n$. 
  Maps with several marked
  edges or vertices are obtained by taking multiple derivatives.
\end{corollary}

We recall that, from the discussion of section \ref{sec:combimaps}, a
labelled map is a map whose half-edges are labelled $\{1,\ldots,2m\}$
where $m$ is the number of edges. Hence by \define{exponential
  generating function for labelled maps} we mean
\begin{equation}
  \label{eq:expgenfun}
  F^{(g)}(t,V) = \sum_{m=0}^{\infty} \frac{t^m}{(2m)!} F^{(g,m)}(V)
\end{equation}
where $F^{(g,m)}(V)$ is the (finite) sum over labelled maps of genus
$g$ with $m$ edges of the product of vertex weights. If we want to
reduce $F^{(g,m)}(V)$ to a sum over unlabeled maps instead, then the
multiplicity of an individual unlabeled map is $(2m)!/\Gamma$, where
$\Gamma$ is its number of automorphisms. The $(2m)!$ cancels the
denominator in (\ref{eq:expgenfun}) leading to an ordinary generating
function, where however the weight $1/\Gamma$ has to be kept. It
differs from the ``true'' generating function for unlabeled maps
where this weight is absent. For rooted maps there is no difference
since $\Gamma=1$ for all of them: we simply refer to \define{the
  generating function for rooted maps} without specifying between
exponential/labelled and ordinary/unlabeled.

$F^{(0)}(t,V)$ is called the \define{planar free energy}. Informally,
it is ``dominant'' in the large $N$ limit. Actually, equation
(\ref{eq:topological}) makes sense as a sum of formal power series (at
a given order in $t$, only a finite number of terms contribute).

\subsection{Gaussian model, Wick theorem}
\label{sec:gaussian}

In order to derive theorem \ref{thm:freetopo}, we first consider the
Gaussian measure
\begin{equation} \label{eq:gaussmeas}
  \left( \frac{N}{2\pi t} \right)^{N^2/2}\, e^{-N \Tr M^2/(2t)}\, dM
\end{equation}
where $dM$ is the Lebesgue (translation-invariant) measure over the
set of Hermitian matrices of size $N$.
It is easily seen that the matrix elements are centered jointly
Gaussian random variables, with covariance
\begin{equation} \label{eq:propag}
  \langle M_{ij} M_{kl} \rangle =
  \delta_{il} \delta_{jk} \frac{t}{N}.
\end{equation}
More generally, the expectation value of the product of arbitrarily
many matrix elements is given via Wick's theorem (generally valid for
any Gaussian measure). This classical result can be stated as follows.
\begin{theorem}[Wick's theorem for matrix integrals] \label{thm:wick}
  The expectation value of the product of an arbitrary number of
  matrix elements is equal to the sum, over all possible pairwise
  matchings of the matrix elements, of the product of pairwise
  covariance.
\end{theorem}
For instance, for $4$ matrix elements we have
\begin{equation}
  \begin{split}
    \langle M_{i_1 j_1} M_{i_2 j_2} M_{i_3 j_3} M_{i_4 j_4}\rangle &=
    \langle M_{i_1 j_1}M_{i_2 j_2}\rangle
    \langle M_{i_3 j_3}M_{i_4 j_4}\rangle \\
    &+ \langle M_{i_1 j_1}M_{i_3 j_3}\rangle
    \langle M_{i_2 j_2}M_{i_4 j_4}\rangle \\
    &+ \langle M_{i_1 j_1}M_{i_4 j_4}\rangle
    \langle M_{i_2 j_2}M_{i_3 j_3}\rangle.
  \end{split}
\end{equation}
Clearly, for an odd number of elements the expectation value vanishes
by parity, while for an even number $2n$ of elements the sum involves
$(2n-1)!! = (2n-1)(2n-3)\cdots 5 \cdot 3 \cdot 1$ terms.

Let us note immediately that these results extend easily to a Gaussian
model of $K$ random Hermitian matrices $M^{(1)},\ldots,M^{(K)}$ of
same size $N$, with measure
\begin{equation}
  \label{eq:multigaussmeas}
  \left( \frac{N^K \det Q}{2^K \pi^K} \right)^{N^2/2}
  \exp \left(- \frac{N}{2} \sum_{a,b=1}^K
    Q_{a b} \Tr M^{(a)} M^{(b)} \right)\, dM^{(1)}\, \cdots\, dM^{(K)}
\end{equation}
where $Q$ is a $K\times K$ real symmetric matrix. The covariance of
two matrix elements is
\begin{equation}
  \langle M_{ij}^{(a)} M_{kl}^{(b)} \rangle =
  \delta_{il} \delta_{jk} \frac{(Q^{-1})_{a b}}{N}.
\end{equation}
and, taking the extra index into account, Wick's theorem still applies,
for instance
\begin{equation}
\begin{split}
  \langle M_{i_1 j_1}^{(a_1)} M_{i_2 j_2}^{(a_2)}
  M_{i_3 j_3}^{(a_3)} M_{i_4 j_4}^{(a_4)} \rangle &=
  \langle M_{i_1 j_1}^{(a_1)} M_{i_2 j_2}^{(a_2)} \rangle
  \langle M_{i_3 j_3}^{(a_3)} M_{i_4 j_4}^{(a_4)} \rangle \\ 
  &+ \langle M_{i_1 j_1}^{(a_1)} M_{i_3 j_3}^{(a_3)} \rangle
  \langle M_{i_2 j_2}^{(a_2)} M_{i_4 j_4}^{(a_4)} \rangle \\
  &+ \langle M_{i_1 j_1}^{(a_1)} M_{i_4 j_4}^{(a_4)} \rangle
  \langle M_{i_2 j_2}^{(a_2)} M_{i_3 j_3}^{(a_3)} \rangle.
\end{split}
\end{equation}

\subsection{Diagrammatics of the one-matrix model: a first approach}
\label{sec:diagrams}

We now return to the partition function (\ref{eq:onemmpfdef}), which
is a formal power series in the variables $v_n$. If
$\mathitbf{k}=(k_n)_{n \geq 1}$ denotes a family of nonnegative
integers with finite support, the coefficient of the monomial
$\mathitbf{v}^{\mathitbf{k}}=\prod_{n=1}^{\infty} v_n^{k_n}$ in
$\Xi_N(t,V)$ reads\footnote{A similar expression appears in
  \cite{Pe88}, albeit with slightly different conventions.}
\begin{equation} \label{eq:onemmcoeff}
  \left[\mathitbf{v}^{\mathitbf{k}} \right]\, \Xi_N(t,V)
  = \left\langle
    \prod_{n=1}^{\infty} \frac{\left(N \Tr M^n\right)^{k_n}}{n^{k_n} k_n!}
  \right\rangle.
\end{equation}
Inside this expression, each trace $\Tr M^n$ can be rewritten as a sum
of product of $n$ elements of $M$, for instance
\begin{equation} \label{eq:trace} \Tr M^3 = \sum_{i,j,k=1}^{N} M_{ij}
  M_{jk} M_{ki}.
\end{equation}
hence (\ref{eq:onemmcoeff}) may itself be rewritten as the expectation
value of a finite linear combination of products of elements of $M$,
to be evaluated via Wick's theorem. 

\begin{figure}[htbp]
  \begin{center}
    \resizebox{\textwidth}{!}{\input{feynman2.pstex_t}}
    \caption{Elements constituting the Feynman diagrams for the
      Hermitian one-matrix model: (a) a leg representing a matrix
      element, (b) a 3-leg vertex representing a term in $\Tr M^3$,
      (c) two paired elements forming an edge.}
    \label{fig:feynman}
  \end{center}
\end{figure}
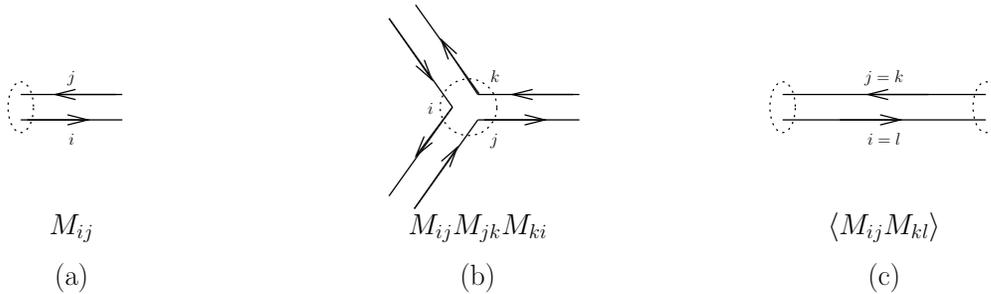

We may represent graphically the factors appearing in this
decomposition as follows (see also Figure \ref{fig:feynman}):
\begin{itemize}
\item[(a)] Each matrix element $M_{ij}$ is represented as a double
  line originating from a point, forming a \define{leg}. The lines are
  oriented in opposite directions (\define{incoming}
  and \define{outgoing}) and ``carry'' respectively an index $i$ and
  $j$.
\item[(b)] Each product of matrix elements appearing in the expansion
  of $\Tr M^n$ is represented as $n$ legs incident to the same point
  forming a \define{vertex}. The legs are cyclically ordered around the
  vertex so that each incoming line is connected to the outgoing
  line of the consecutive leg, and carries the same index. This
  directly translates the pattern of indices obtained when writing a
  trace as product of matrix elements. For instance, for $n=3$ the
  decomposition (\ref{eq:trace}) yields the vertex shown on Figure
  \ref{fig:feynman}(b), where each index $i$, $j$, or $k$ takes $N$
  possible values.
\item[(c)] Wick's theorem states that the expectation value of a
  product of matrix elements is obtained by matching them pairwise in
  all possible manners, and taking the corresponding product of
  covariances. A pair of matched elements is represented by linking
  the corresponding legs, forming an \define{edge}. More
  precisely, the incoming line of the one leg is connected to the
  outgoing line of the other leg, and carries the same index. This
  translates relation (\ref{eq:propag}): if the connected lines do not
  carry the same index, then the covariance vanishes hence the
  matching does not contribute to the expectation value.
\end{itemize}

Globally, the factors appearing in (\ref{eq:onemmcoeff}) form a
collection of $k=\sum k_n$ vertices, consisting of $k_n$ $n$-leg
vertices for all $n \geq 1$. The expectation value is obtained by
matching the $\sum n k_n$ legs, i.e merging them pairwise into edges,
in all possible manners. Hence, the quantity (\ref{eq:onemmcoeff}) is
expressed as a sum over all diagrams built out of these vertices and
edges, which are sometimes called \define{fatgraphs} or \define{ribbon
  graphs} in the literature.  By the rules discussed above, the index
lines that are merged together must carry the same index, and they
form closed oriented cycles. There is clearly a finite number of
diagrams, since there are $(\sum n k_n)!!$ ways to merge the legs (in
particular, if the number of legs is odd, there are no such diagrams,
and correspondingly the expectation value vanishes). It remains to
determine what is the contribution of an individual diagram. Because
of relation (\ref{eq:propag}), each of the $m=\sum n k_n/2$ edges
produces a factor $1/\lambda=t/N$. Hence all diagrams will have the
same contribution $t^{m} N^{k-m} / \prod_{n=1}^{\infty} n^{k_n} k_n!$,
taking into account the extra factors present in
(\ref{eq:onemmcoeff}). Evaluating this matrix integral amounts to
counting the number of such diagrams.

\begin{figure}[htbp]
  \begin{center}
    \includegraphics[width=\textwidth]{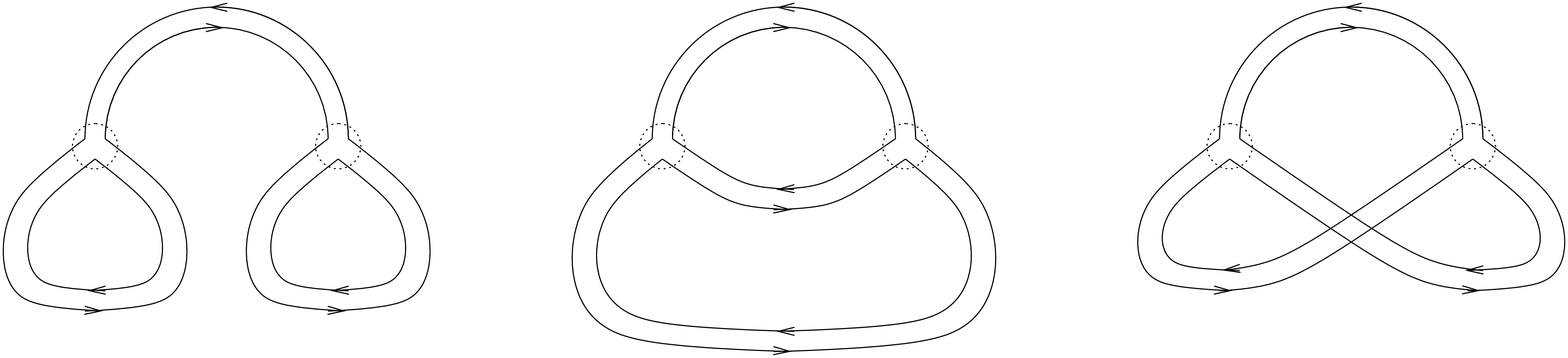}
    \caption{The three possible types of diagram appearing in the
      expansion of \mbox{$\langle \Tr M^3 \cdot \Tr M^3 \rangle$.}}
    \label{fig:fatgraphs}
  \end{center}
\end{figure}

However, the sum involves many ``equivalent'' diagrams, i.e diagrams
which differ only by the choice of line indices or which have the same
``shape''. For instance, Figure \ref{fig:fatgraphs} displays the three
possible types of diagrams obtained when expanding $\langle \Tr M^3
\cdot \Tr M^3 \rangle$ (in this example, all three diagrams are
connected but this is not true in general). Clearly, we may forget
about the line indices by counting each index-less diagram with a
multiplicity $N^{\ell}$, where $\ell$ is the number of cycles of index
lines. In our example, we have $\ell=3$ for the first two diagrams
while $\ell=1$ for the third. Evaluating the shape multiplicity is
slightly more subtle, and we leave its general discussion to the next
subsection. It is not difficult to do the computation for the diagrams
of Figure \ref{fig:fatgraphs}, which yields the respective shape
multiplicities 9, 3 and 3, and gathering the various factors we arrive
at
\begin{equation}
  \langle \Tr M^3 \cdot \Tr M^3 \rangle = (9 N^3 + 3 N^3 + 3 N) (t/N)^3
  = \left( 12 + \frac{3}{N^2} \right) t^3.
\end{equation}
Generally, as mentioned above, the coefficient (\ref{eq:onemmcoeff})
will correspond to a sum over all (not necessarily connected) diagrams
made out of $k_n$ vertices with $n$ legs for all $n$. The partition
function $\Xi_N(t,V)$ is then obtained by summing over all $k_n$'s,
attaching a weight $v_n$ per $n$-leg vertex, leading to the generating
function for all diagrams (possibly empty or disconnected). Then
$F_N(t,V)=\log \Xi_N(t,V)$ is the generating function for connected
diagrams, as it is well-known. In $\Xi_N(t,V)$ as well as in
$F_N(t,V)$, the exponent of $N$ in the contribution of a diagram is
equal to the number of vertices minus the number of edges plus the
number of index lines.

\begin{figure}[htbp]
  \begin{center}
    \includegraphics[width=\textwidth]{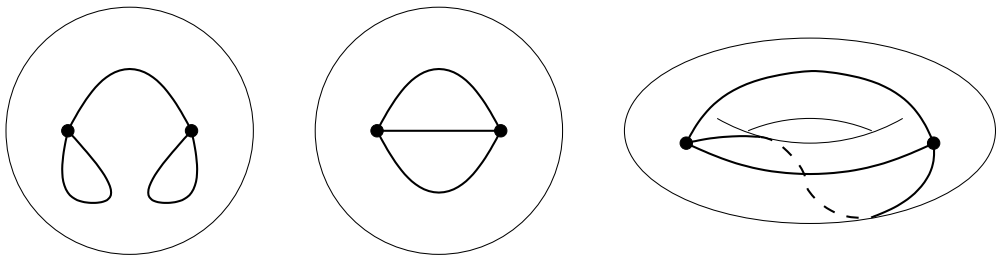}
    \caption{The maps corresponding to the diagrams of figure
      \ref{fig:fatgraphs}: the first two have genus 0, the third genus
      1.}
    \label{fig:fatgraphs2}
  \end{center}
\end{figure}

At this stage, it might be rather clear that our (connected) diagrams
are nothing but maps in disguise (see figure
\ref{fig:fatgraphs2}). Indeed they are graphs endowed with a cyclic
order of half-edges (legs) around the vertices, which is a
characterization of maps as discussed in section
\ref{sec:mapdefs}. The cycles of index lines correspond to faces,
hence the exponent of $N$ in the contribution of a diagram to
$F_N(t,V)$ is equal to the Euler characteristic of the corresponding
map. This essentially establishes theorem \ref{thm:freetopo}.

\subsection{One-matrix model and combinatorial maps}
\label{sec:matrixcombi}

In this section, we revisit the calculation done above in a more
formal manner. The purpose is to show that it naturally involves
labelled combinatorial maps, i.e pairs of permutations, as defined in
section \ref{sec:combimaps}.

We start from a ``classical'' formula for enumerating permutations
with prescribed cycle lengths. Let $\mathcal{S}_p$ denote the set of
permutations of $\{1,\ldots,p\}$ and $c_n(\sigma)$ denote the number
of $n$-cycles in the permutation $\sigma$, $c(\sigma)$ being the total
number of cycles. Then the numbers of permutations $\sigma$ in
$\mathcal{S}_p$ with prescribed values of $c_n(\sigma)$ for all $n
\geq 1$ are encoded into the exponential multivariate generating
function
\begin{equation}
  \label{eq:permexp}
  \exp \left( \sum_{n=1}^{\infty} \frac{A_n}{n} \right) =
  \sum_{p=0}^{\infty}  \frac{1}{p!} \sum_{\sigma \in \mathcal{S}_p}
  \prod_{n=1}^{\infty} A_n^{c_n(\sigma)}.
\end{equation}
Here $(A_n)_{n \geq 1}$ is a family of formal variables. Establishing
this formula is a simple exercise in combinatorics \cite{AnaComb}: it
simply translates the decomposition of a permutation into cycles.  We
recover $\Xi_N(t,V)$ of (\ref{eq:perturb}) on the left-hand side by
the substitution
\begin{equation}
  \label{eq:expsubs}
  A_n = N v_n \Tr M^n
\end{equation}
and taking the expectation value over the Gaussian random matrix
$M$. By this substitution, the $\sigma$ term on the right hand side
is, up to a factor independent of $M$, equal to a product of traces
which we may rewrite as
\begin{equation} \label{eq:traceperm}
  \prod_{n=1}^{\infty} (\Tr M^n)^{c_n(\sigma)} =
  \sum_{(i_1,\ldots,i_p) \in \{1,\ldots,N\}^p} \prod_{q=1}^{p} M_{i_q\, i_{\sigma(q)}}.
\end{equation}
Wick's theorem and relation (\ref{eq:propag}) yield 
\begin{equation} \label{eq:permwick}
  \left\langle \prod_{q=1}^{p} M_{i_q\, i_{\sigma(q)}} \right\rangle = \left( \frac{t}{N} \right)^{p/2} \sum_{\alpha \in \mathcal{I}_p}
  \prod_{q=1}^{p} \delta_{i_q\, i_{\sigma(\alpha(q))}}
\end{equation}
where $\mathcal{I}_p \subset \mathcal{S}_p$ is the set of involutions
without fixed point (aka pairwise matchings) of $\{1,\ldots,p\}$, which
is empty for $p$ odd. We then observe that the product on the right
hand side of (\ref{eq:permwick}) is equal to 1 if the index $i_q$ is
constant over the cycles of the permutation $\sigma \circ \alpha$,
otherwise it is 0. Therefore, when summing over all values of
$(i_1,\ldots,i_p)$, we find
\begin{equation}
  \label{eq:tracewick}
  \left\langle \prod_{n=1}^{\infty} (\Tr M^n)^{c_n(\sigma)}
  \right\rangle = \left( \frac{t}{N} \right)^{p/2}
  \sum_{\alpha \in \mathcal{I}_{p}} N^{c(\sigma \circ \alpha)}.
\end{equation}
Plugging into (\ref{eq:permexp}) and (\ref{eq:expsubs}) and writing
$p=2m=2c(\alpha)$, we arrive at
\begin{equation}
  \Xi_N(t,V) = \sum_{m=0}^{\infty} \frac{t^m}{(2m)!} \sum_{(\sigma,\alpha)
    \in \mathcal{S}_{2m} \times \mathcal{I}_{2m}}
  N^{c(\sigma) - c(\alpha) + c(\sigma \circ\alpha)} \prod_{n=1}^{\infty} v_n^{c_n(\sigma)}.
\end{equation}
We are very close to recognizing a sum over combinatorial maps, with
the Euler characteristic (\ref{eq:eulerperm}) appearing as the
exponent of $N$, but we lack the requirement that $\sigma$ and
$\alpha$ generate a transitive subgroup. Again, this is implemented by
taking the logarithm (in combinatorial terms \cite{AnaComb},
$\mathcal{S}_{2m} \times \mathcal{I}_{2m}$ is equal to the labelled
set construction applied to the class of combinatorial maps -- as seen
by decomposing $\{1,\ldots,2m\}$ into orbits -- and all parameters are
inherited, hence $\Xi_N(t,V)$ is the exponential of the generating
function for combinatorial maps), leading to
\begin{equation}
  F_N(t,V) = \log \Xi_N(t,V)
  = \sum_{m=0}^{\infty} \frac{t^{m}}{(2m)!} \sum_{(\sigma,\alpha)
    \in \mathcal{M}_m} N^{\chi(\sigma,\alpha)} \prod_{n=1}^{\infty} v_n^{c_n(\sigma)}
\end{equation}
where $\mathcal{M}_m$ is the set of labelled combinatorial maps with
$m$ edges. Upon regrouping the maps according to their genus, this
yields relation (\ref{eq:expgenfun}) and formally establishes
theorem \ref{thm:freetopo}.

\subsection{Generalization to multi-matrix models}
\label{sec:multimat}

Let us now briefly discuss multi-matrix models. Informally, we
consider a ``perturbation'' of the multi-matrix Gaussian model
(\ref{eq:multigaussmeas}) by a $U(N)$-invariant potential $N \Tr V
(M^{(1)},M^{(2)},\ldots,M^{(K)})$, where $V(x_1,x_2,\ldots,x_K)$ is a
``polynomial'' in $K$ non-commutative variables. Actually, $V$ shall
be viewed as a formal linear combination of monomials in $p$
non-commutative variables $x_1,x_2,\ldots,x_K$ namely
\begin{equation}
  \label{eq:vmulti}
  V(x_1,x_2,\ldots,x_K) = \sum_{n=0}^{\infty}
  \sum_{(a_1,\ldots,a_n) \in \{1,\ldots,K\}^n} \frac{v(a_1,\ldots,a_n)}{n} x_{a_1} \cdots x_{a_n}
\end{equation}
where the $v(a_1,\ldots,a_n)$ are formal variables with the
identification $v(a_2,\ldots,a_n,a_1)=v(a_1,a_2,\ldots,a_n)$ (since
the trace is invariant by cyclic shifts). The partition function is
defined as the expectation value of $e^{N \Tr
  V(M^{(1)},M^{(2)},\ldots,M^{(K)})}$ under the Gaussian measure
(\ref{eq:multigaussmeas}), and the free energy as its logarithm.  By a
simple extension of the arguments of section \ref{sec:diagrams} (using
Wick's theorem for multi-matrix integrals), the free energy may be
written as a sum over maps, whose half-edges now carry one of $K$
``colours'' corresponding to the extra index $a=1,\ldots,K$. The
formal variable $v(a_1,\ldots,a_n)$ is the weight for vertices around
which the half-edge colors are $(a_1,\ldots,a_n)$ in cyclic order, and
$Q^{-1}_{ab}$ is the weight per edge whose halves are colored $a$ and
$b$. Furthermore the topological expansion (\ref{eq:topological})
still holds.

In the particular case
\begin{equation} \label{eq:Vmulticol} V (x_1,x_2,\ldots,x_K) =
  \sum_{n=1}^{\infty} \sum_{a=1}^K \frac{v_n^{(a)}}{n} x_{a}^n,
\end{equation}
all legs incident to a same vertex carry the same colour. The free
energy yields the generating function for (labelled) maps of a given
genus whose vertices are colored in $K$ colours, with a weight
$v_n^{(a)}$ per vertex with colour $a$ and degree $n$, and a weight
$Q^{-1}_{a,b}$ per edge linking of vertex of colour $a$ to a vertex of
colour $b$.

Instances of these models appears in several other chapters. Those
corresponding to the form (\ref{eq:Vmulticol}) include the chain
matrix model in chapter \Orantin, the Ising and Potts models in
chapter \Kostov.  Models of the form (\ref{eq:vmulti}) but not
(\ref{eq:Vmulticol}) include the complex matrix model in chapter
\ZinnZuber (for the diagrammatic expansion, $M$ and $M^{\dagger}$ may
be treated as two independent Hermitian matrices, which are
represented with outgoing and incoming arrows), the $O(n)$,
six-vertex and SOS/ADE models in chapter \Kostov.  We particularly
emphasize the two-matrix model with free energy
\begin{equation}
  \label{eq:twomatfree}
  \log \frac{\int \, \exp \left(
      - \frac{N}{t} \Tr M_1 M_2 + \sum_n \frac{N
        v_n^{(1)}}{n} \Tr M_1^n + \sum_n \frac{N
        v_n^{(2)}}{n} \Tr M_2^n \right) \, dM_1\, dM_2}{\int\,
    \exp \left( \frac{N}{t} \Tr M_1 M_2 \right) dM_1\, dM_2\,}
\end{equation}
which yields generating functions for bipartite maps with a weight per
vertex depending on degree and colour. It is the most natural
generalization of the counting problem addressed with the one-matrix
model (recovered by setting $v^{(2)}_n=\delta_{n,2}/t$).

Let us finally mention that the diagrammatic expansion for real
symmetric matrices corresponds to maps on unoriented surfaces.

\sect{The vertex degree distribution of planar maps}
\label{sec:onemmsol}


This section is devoted to the enumeration of rooted planar maps with
a prescribed vertex degree distribution, i.e with a given number of
vertices of each degree. This is equivalent to deriving the generating
function for rooted planar maps with a weight $t$ per edge and, for
all $n \geq 1$, a weight $v_n$ per vertex of degree $n$. By theorem
\ref{thm:freetopo} and its corollary, this in turn amounts to
computing the derivative with respect to $t$ of the planar free energy
of the one-matrix model.

Let us first state the result, in the form given in \cite{census}.  A
different form was obtained independently in \cite{BeCa94} (without
matrices), and a check of their agreement can be found in
\cite{BoJe06}.

\begin{theorem}
  \label{thm:census}
  Let $R,S$ be the formal power series in $t$ and $(v_n)_{n \geq 1}$ satisfying
  \begin{equation} \label{eq:rsthm}
    \begin{split}
      R &= t + t \sum_{n=1}^{\infty} v_n \sum_{j=1}^{\lfloor
        \frac{n}{2} \rfloor}
      \frac{(n-1)!}{j!(j-1)!(n-2j)!} R^j S^{n-2j} \\
      S &= t \sum_{n=1}^{\infty} v_n \sum_{j=0}^{\lfloor \frac{n-1}{2}
        \rfloor} \frac{(n-1)!}{(j!)^2 (n-2j-1)!} R^j S^{n-2j-1}.
    \end{split}
  \end{equation}  
  Then the generating function of rooted planar maps with a weight $t$
  per edge and, for all $n$, a weight $v_n$ per vertex of degree $n$
  is given by
  \begin{equation}
    \label{eq:rootthm}
    E^{(0)} = \frac{1}{t}
    \left( R + S^2 - \sum_{n=1}^{\infty} v_n \sum_{j=2}^{\lfloor \frac{n+2}{2} \rfloor}
      \frac{(2n-3j+2)(n-1)!}{j!(j-2)!(n-2j+2)!} R^j S^{n-2j+2} - t \right).
  \end{equation}
\end{theorem}

\begin{remark} Clearly, $R,S$ are uniquely determined from the
  requirement that $R=S=0$ for $t=0$. They have a direct combinatorial
  interpretation: $R$ is the generating function for planar maps with
  two distinguished vertices of degree 1 (without weights), $S$ is the
  generating function for planar maps with one distinguished vertex
  (without weight) of degree 1 and one distinguished face.
\end{remark}

Let us mention that the planar free energy $F^{(0)}$ itself has a more
complicated expression, involving logarithms.

This section is organized as follows. In subsection (\ref{sec:master})
we discuss the main equation describing the planar limit. In
subsection (\ref{sec:onecut}) we explain how to solve this equation
and derive theorem \ref{thm:census}. Finally in subsection
(\ref{sec:examap}) we present a few instances with explicit counting
formulas.

\subsection{Saddle-point, loop, Tutte's equations}
\label{sec:master}

Our goal is to ``solve'' the one-matrix model in the large $N$ limit,
i.e extract the genus 0 contribution in (\ref{eq:topological}). This
problem has already been approached several times in this book
(particularly in chapters \CicutaMolinari and \Orantin), let us recall
what the ``master'' equation is. It is an equation is for a quantity
called the resolvent in the context of matrix integrals. For our
purposes, it is nothing but a generating function for rooted maps
involving an extra variable attached to the degree of the root vertex.

More precisely, we define here the planar resolvent as
\begin{equation}
  \label{eq:resolvdef}
  W(z)=\sum_{n=0}^{\infty} \frac{W_n}{z^{n+1}}=\sum_{n=0}^{\infty}
  \frac{n}{z^{n+1}} \frac{\partial F_n^{(0)}}{\partial v_n}
\end{equation}
where $z$ is a new formal variable and $W_n$ is the generating
function for rooted (i.e with a distinguished half-edge) planar maps
whose root vertex (i.e the vertex incident to the root) has degree
$n$, with a weight $t$ per edge and, for all $n$, a weight $v_n$ per
non-root vertex. By convention we set $W_0=1$. In comparison with the
corollary of theorem \ref{thm:freetopo} we do not attach a weight to
the root vertex.

Then, the planar resolvent satisfies the master equation
\begin{equation} \label{eq:quadratresolv}
  W(z)^2 - \left( \frac{z}{t} - V'(z) \right) W(z) + P(z) = 0
\end{equation}
which is a quadratic equation for $W(z)$ immediately solved into
\begin{equation}
  \label{eq:quadratsol}
  W(z) = \frac{1}{2} \left( \frac{z}{t} - V'(z)
    \pm \sqrt{ \left(\frac{z}{t} - V'(z) \right)^2 - 4 P(z) }
  \right).
\end{equation}
At this stage $P(z)$ is still an unknown quantity but all derivations
of (\ref{eq:quadratresolv}) show that, unlike $W(z)$, \emph{$P(z)$
  contains only non-negative powers of $z$}. Actually, if $V(z)$ is a
polynomial of degree $d$ (i.e we set $v_n=0$ for $n>d$), then $P(z)$
is a polynomial of degree $d-2$.  These remarks are instrumental in
solving the equation, as discussed in the next subsection. Let us
first briefly review some methods for deriving the master equation.

\noindent {\bf Saddle-point approximation.}
The saddle-point approximation is the original ``physical'' method
used in \cite{BIPZ}. It consists in treating the partition function
(\ref{eq:perturb}) as a ``genuine'' matrix integral and extracting its
analytical large $N$ asymptotics. This is done classically by reducing
to an integral over the eigenvalues, then determining the dominant
eigenvalue distribution. See chapter \CicutaMolinari, sections 1 and
2, for a general discussion of this method. Equation
(\CicutaMolinari.2.6) is nothing but equation (\ref{eq:quadratsol}) in
different notations: the quantities denoted by $F(z)$, $Q(z)$ and
$V'(z)$ in chapter \CicutaMolinari correspond respectively to $W(z)$,
$P(z)$ and $\frac{z}{t} - V'(z)$ here.

\noindent {\bf Loop equations.}
Loop equations correspond to the Schwinger--Dyson equations of quantum
field theory applied in the context of matrix models
\cite{Wad81,Mig83}, see chapter \Orantin for a general discussion and
application in the context of the one-matrix model. An interesting
feature of loop equations is that they provide an easier access to
higher genus contributions than the saddle-point approximation.
However we are here interested in the planar case for which they are
essentially equivalent. Again, equation (\Orantin.4.1) is
(\ref{eq:quadratresolv}) in different notations: the quantities
denoted by $W_1^{(0)}(x)$, $P^{(0)}(z)$ and $V'(z)$ in chapter
\Orantin correspond respectively to $W(z)$, $P(z)$ and $\frac{z}{t} -
V'(z)$ here.

\noindent {\bf Tutte's recursive decomposition.}
Tutte's original approach consists in recursively decomposing rooted
maps by ``removing'' (contracting or deleting) the root edge. It
translates into an equation determining their generating function,
upon introducing an extra ``catalytic'' variable in order to make the
decomposition bijective. It is now recognized that Tutte's equations
are essentially equivalent to loop equations \cite{Eyn06} despite
their very different origin.

Let us explain the recursive decomposition in our setting
\cite{Tut68,BoJe06}. We consider a rooted planar map whose root
degree (i.e the degree of the root vertex) is $n$, and we decompose it
as follows.
\begin{itemize}
\item If the root edge is a loop (i.e connects the root vertex to
  itself), then it naturally ``splits'' the map into two parts, which
  may be viewed as two rooted planar maps. If there are $i$ half-edges
  incident to the root vertex on one side (excluding those of the
  loop), then there are $n-2-i$ on the other side. These are the respective
  root degrees of the corresponding maps.
\item If the root edge is not a loop, then we contract it (and we may
  canonically pick a new root). If $m$ denotes the degree of the
  other vertex incident to the root edge in the original map, then
  the root degree of the contracted map is $n+m-2$.
\end{itemize}
This decomposition is clearly reversible. Taking into account the
weights, it leads to the equation
\begin{equation}
  W_n = t \sum_{i=0}^{n-2} W_i W_{n-2-j} +
  t \sum_{m=1}^{\infty} v_m W_{n+m-2}
\end{equation}
valid for all $n \geq 1$ with the convention $W_0=1$. Equation
(\ref{eq:quadratresolv}) is deduced using (\ref{eq:resolvdef}), in
particular $P(z)$ is given by
\begin{equation}
  P(z) = - \sum_{n=0}^{\infty} \left( \sum_{m=n+2}^{\infty} v_m
  W_{m-2-n} \right) z^n.
\end{equation}

\subsection{One-cut solution}
\label{sec:onecut}

We now turn to the solution of equation (\ref{eq:quadratresolv}). In
the context of matrix models, it gives the ``one-cut
solution'' discussed for instance in chapter \CicutaMolinari. Here we
concentrate on expressing it in combinatorial form.  Let us first
suppose that $V(z)$ (hence $P(z)$) is a polynomial in $z$. The one-cut
solution is obtained by assuming that the polynomial
$\Delta(z)=\left(z/t - V'(z) \right)^2 - 4 P(z)$ appearing under the
square root in (\ref{eq:quadratsol}) has exactly two simple zeroes,
say in $a$ and $b$, and only double zeroes elsewhere. This leads to
\begin{equation} \label{eq:omegaonecut}
  W(z) = \frac{1}{2} \left( \frac{z}{t} - V'(z) +
     G(z) \sqrt{(z-a)(z-b)} \right)
\end{equation}
where $G(z)$ is a polynomial. This assumption is physically justified
in the saddle-point picture by saying that the dominant eigenvalue
distribution has a support made of a single interval $[a,b]$
(corresponding to the cut of $W(z)$), as a perturbation of Wigner's
semi-circle distribution. Alternatively, a rigorous proof comes via
Brown's lemma \cite{Bro65}, which translates the fact that $W(z)$
hence $\sqrt{\Delta(z)}$ are power series without fractional powers,
we refer to \cite[section 10]{BoJe06} for details in the current
context. $\sqrt{(z-a)(z-b)}$ must be understood as a Laurent series in
$1/z$, i.e $\sqrt{(z-a)(z-b)}=z+\text{(lower powers in $z$)}$.

Now, it turns out that $a,b$ and $G(z)$ in (\ref{eq:omegaonecut}) may
be fully determined from the mere condition that $W(z) = 1/z +
\text{(lower powers in $z$)}$. Indeed, let us rewrite
(\ref{eq:omegaonecut}) as
\begin{equation}
  \label{eq:omegadivid}
  \frac{W(z)}{\sqrt{(z-a)(z-b)}} =  \frac{1}{2} \left( 
    \frac{\frac{z}{t} - V'(z)}{\sqrt{(z-a)(z-b)}} + G(z) \right).
\end{equation}
Then, we first extract the coefficients of $z^{-1}$ and $z^{-2}$ on
both sides. On the left hand side, we obtain respectively $0$ and $1$
by the above condition. On the right hand side, $G(z)$ does not
contribute since it is a polynomial in $z$. Therefore we arrive at
\begin{equation}
  \label{eq:omegadivcoeffs}
    \left. \frac{\frac{z}{t} -
        V'(z)}{\sqrt{(z-a)(z-b)}} \right|_{z^{-1}}=0 \qquad
    \left. \frac{\frac{z}{t} -
      V'(z)}{\sqrt{(z-a)(z-b)}} \right|_{z^{-2}}=2.
\end{equation}
These equations determine $a$ and $b$ in terms of $t$ and $V(z)$. The
coefficients may be extracted via a contour integration around
$z=\infty$, but a nicer form is obtained by performing a change of
variable $z \to u$ given by
\begin{equation}
  \label{eq:jouk}
  z = u + S + \frac{R}{u}
\end{equation}
also known as Joukowsky's transform. $S$ and $R$ are chosen such that
$(z-a)(z-b)$ becomes a perfect square namely
\begin{equation}
  \label{eq:srab}
  S = \frac{a+b}{2} \qquad R = \frac{(b-a)^2}{16} \qquad
  (z-a)(z-b)=\left(u - \frac{R}{u} \right)^2.
\end{equation}
Then, by this change of variable, relations (\ref{eq:omegadivcoeffs})
yield
\begin{equation} \label{eq:rsrecur}
    S = t \left. V'\left(u + S + \frac{R}{u}\right) \right|_{u^0} \qquad
    R = t + t \left. V'\left(u + S + \frac{R}{u}\right)
    \right|_{u^{-1}}.
\end{equation}
Upon expanding $V'(z)=\sum v_n z^n$, then extracting the respective
coefficients of $u^0$ and $u^{-1}$ in $(u+S+R/u)^k$ via the
multinomial formula, we obtain the equations (\ref{eq:rsthm}).

Extracting further coefficients $z^{-3},z^{-4},\ldots$ in
(\ref{eq:omegadivid}) , we may determine step by step the first few
$W_n$. In particular, $W_2$ is, up to a factor, the generating
function $E^{(0)}$ for rooted planar maps (without condition on the
root vertex), since marking a bivalent vertex is tantamount to marking
an edge. A slightly tedious computation yields
\begin{equation}
  \label{eq:onemmsol}
  E^{(0)} = \frac{R + S^2 - \left. V'\left(u + S +
        \frac{R}{u}\right) \right|_{u^{-3}} - 2 S \left.
      V'\left(u + S + \frac{R}{u}\right) \right|_{u^{-2}} - t}{t}.
\end{equation}
which can then be put into the form (\ref{eq:rootthm}). This
establishes theorem \ref{thm:census}, upon noting the restriction that
$V(z)$ is a polynomial may now be ``lifted'': at a given order in $t$,
the coefficient of $E^{(0)}$ and the corresponding sum over maps both
depend on finitely many $v_n$, therefore by suitably truncating $V(z)$
we may establish their equality.

\subsection{Examples}
\label{sec:examap}

\noindent {\bf Tetravalent maps.} 
In the case of a quartic potential $V(x)=x^4/4$, the diagrammatic
expansion involves only vertices of degree 4, forming 4-regular or
\define{tetravalent} maps. Equations (\ref{eq:rsthm}) and
(\ref{eq:rootthm}) reduce to
\begin{equation}
  \label{eq:tetra}
  E^{(0)}_4 = \frac{R - R^3}{t} \qquad S=0 \qquad R = t + 3 t R^2.
\end{equation}
$R$ is given by a particularly simple quadratic equation solved as
\begin{equation}
  \label{eq:catalan}
  R = \frac{1 - \sqrt{1 - 12 t^2}}{6 t} =
  \sum_{k=0}^{\infty} \frac{1}{k+1} \binom{2k}{k}\, 3^k\, t^{2k+1}
\end{equation}
where we recognize the celebrated Catalan numbers. Substituting into
$E^{(0)}_4$ we obtain the series expansion
\begin{equation}
  E^{(0)}_4 = \sum_{k=0}^{\infty} \frac{2 (2k)!}{k!(k+2)!}\, 3^k\, t^{2k}
\end{equation}
where we identify the number of rooted planar tetravalent maps with
$2k$ edges (hence $k$ vertices). This is the same number as that of
(general) rooted planar maps with $k$ edges \cite{Tut63}, as seen by
Tutte's equivalence, and that of rooted planar quadrangulations (i.e
maps with only faces of degree 4) with $k$ faces, as seen by duality.

\noindent {\bf Trivalent maps.}
In the case of a cubic potential $V(x)=x^3/3$, the diagrammatic
expansion involves only vertices of degree 3, forming 3-regular or
\define{trivalent} maps. Equations (\ref{eq:rsthm}) and
(\ref{eq:rootthm}) reduce to
\begin{equation}
  E^{(0)}_3 =\frac{R+S^2-2R^2S-t}{t} \qquad S = t (S^2+2R)
 \qquad R = t + 2 t R S
\end{equation}
and we may eliminate $R$, yielding a cubic equation for $S$ namely
\begin{equation}
  t^3 = \frac{t S(1-t S)(1-2 t S)}{2}.
\end{equation}
(hence $t S$ will be a power series in $t^3$). Substituting into
$E^{(0)}_3$, we may use the Lagrange inversion formula to compute
explicitly its expansion as
\begin{equation} \label{eq:triplan}
  E^{(0)}_3 = \sum_{k=0}^{\infty} \frac{2^{2k+1} (3k)!!}{(k+2)! k!!} t^{3k}.
\end{equation}
where we recognize the number of rooted planar trivalent maps with
$3k$ edges (hence $2k$ vertices) \cite{MuNeSc70}.

\noindent {\bf Eulerian maps.}
We now consider the case of a general even potential ($v_n=0$ for $n$
odd). This corresponds to counting maps with vertices of even degree,
which are called \define{Eulerian} (these maps admit a Eulerian path,
i.e a path visiting each edge exactly once). A drastic simplification
occurs in (\ref{eq:rsthm}), namely that $S=0$ and $R$ is given by
\begin{equation}
  R = t + t \sum_{n=1}^{\infty} v_{2n} \binom{2n-1}{n} R^n.
\end{equation}
Hence the generating function $E^{(0)}$ for rooted planar Eulerian
maps depends on a single function $R$ satisfying an algebraic
equation. This paves the way to an application of the Lagrange
inversion formula, allowing to compute the general term in the series
expansion of $E^{(0)}$ namely the number of rooted planar Eulerian
maps having a prescribed number $k_n$ of vertices of degree $2n$ for
all $n$, given by
\begin{equation}
  \label{eq:eulermaps}
  2 \frac{\left( \sum_{n=1}^{\infty} n k_n \right)!}{\left( \sum_{n=1}^{\infty}
      (n-1) k_n + 2 \right)!} \prod_{k=1}^{\infty} \frac{1}{k_n!}
  \binom{2k-1}{k}^{k_n}.
\end{equation}
This formula was first derived combinatorially by Tutte \cite{Tut62c}.

\sect{From matrix models to bijections}
\label{sec:bij}

To conclude this chapter, we move a little away from matrices and
explain how to rederive the enumeration result of theorem
\ref{thm:census} through a bijective approach. Such an approach consists in
counting objects (here maps) by transforming them into other objects
easier to enumerate. We have already encountered several bijections in
this chapter: between maps and some pairs of permutations, between
fatgraphs and maps, in Tutte's recursive decomposition. However they
do not directly yield an enumeration formula, as a non-bijective step
is needed.

The ``easier'' objects we shall look for are \define{rooted plane
  trees} (which we may view as rooted planar maps with one face). This
might not come as a surprise ever since the appearance of Catalan
numbers at equation (\ref{eq:catalan}). In general, trees are indeed
easy to enumerate by recursive decomposition: removing the root cuts
the tree into subtrees (forming an ordered sequence due to planarity)
and this is often immediately translated into an algebraic equation
for their generating function. Here, we will perform the inverse
translation: we will \emph{construct} the trees corresponding to a
given equation, namely (\ref{eq:rsthm}) or equivalently
(\ref{eq:rsrecur}).

\begin{figure}[htpb]
  \centering
  \includegraphics[width=\textwidth]{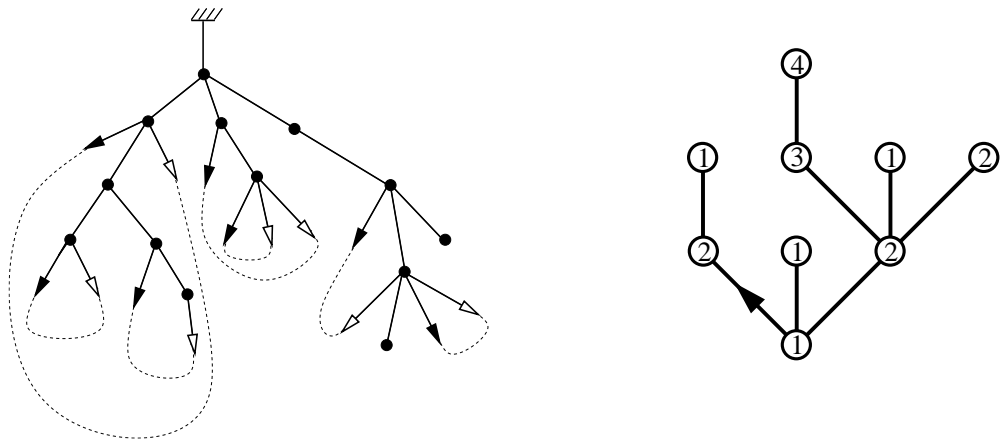}
  \caption{Left: a $\mathcal{S}$-tree, together with the canonical
    matching of black and white leaves (dashed lines between
    arrows). Right: a well-labeled tree.}
  \label{fig:blossom_match}
\end{figure}

Let us indeed define two classes of rooted trees $\mathcal{R}$ and
$\mathcal{S}$ recursively as follows. A $\mathcal{R}$-tree (i.e a tree
in $\mathcal{R}$) is either reduced to a ``white'' leaf, or it
consists of a root vertex to which are attached a sequence of subtrees
that can be $\mathcal{R}$-trees, $\mathcal{S}$-trees or single
``black'' leaves, with the condition that the number of such black
leaves is equal to the number of $\mathcal{R}$-subtrees minus one.  A
$\mathcal{S}$-tree consists of a root vertex to which are attached a
sequence of the same possible subtrees, with the condition that the
number of black leaves is equal to the number of
$\mathcal{R}$-subtrees. It is straightforward to check that this is a
well-defined recursive construction, which translates into the
equations (\ref{eq:rsthm}) or (\ref{eq:rsrecur}) for the corresponding
generating functions $R$ and $S$, provided that we attach a weight $t$
per vertex or white leaf, and a weight $v_n$ per vertex with $n-1$
subtrees. Figure \ref{fig:blossom_match} displays a tree in
$\mathcal{S}$.

We may then wonder how such trees are related to maps. A natural idea
is to ``match'' the black and white leaves together, creating new
edges. Consider for instance a $\mathcal{S}$-tree: it has the same
number of black and white leaves, and given an orientation there is a
``canonical'' matching procedure, see again figure
\ref{fig:blossom_match}. This creates a planar map out of a
$\mathcal{S}$-tree, and we further observe that the vertex degrees are
preserved. It is possible to show that this defines a one-to-one
correspondence between $\mathcal{S}$ and the second class of maps
mentioned in the remark below theorem \ref{thm:census}, see
\cite{census} for details. We proceed similarly with $\mathcal{R}$, and
then a few more steps allow to establish bijectively theorem
\ref{thm:census}. The knowledge of the matrix model solution was
instrumental in ``guessing'' the suitable family of trees, which
encompasses the one found in \cite{Sch97} based on Tutte's formula
(\ref{eq:eulermaps}). The present construction was further extended to
bipartite maps corresponding to the two-matrix model
(\ref{eq:twomatfree}) \cite{BMS02}, and maps corresponding to a
chain-matrix model \cite{HObipar}.

The bijective approach has a great virtue. It was indeed realized that
it is intimately connected with the ``geodesic'' distance in maps
\cite{geod}. Actually, there is another ``dual'' family of bijections
with so-called well-labeled trees or mobiles
\cite{CV81,MS01,mobiles,fomap} for which the connection is even more
apparent. In the simplest instance \cite{CV81}, a well-labeled tree is
a rooted plane tree whose vertices carry a positive integer label, in
such a way that labels on adjacent vertices differ by at most 1 (see
again figure \ref{fig:blossom_match} for an example). It encodes
bijectively a rooted quadrangulation (i.e a map whose faces have
degree 4), a vertex with label $\ell$ in the tree corresponding to a
vertex at \emph{distance} $\ell$ from the root vertex in the
quadrangulation \cite{MS01} (where the distance is the graph distance,
i.e the minimal number of consecutive edges connecting two
vertices). It is easily seen that the generating function for
well-labeled trees with root label $\ell \geq 1$ satisfies
\begin{equation}
  \label{eq:wlrecur}
  R_{\ell} = t + t R_{\ell} \left( R_{\ell-1} + R_{\ell} + R_{\ell+1} \right)
\end{equation}
together with the boundary condition $R_0=0$, $t$ being a weight per
edge or vertex. Through the bijection, $R_{\ell}$ yields the
generating function for quadrangulations with two marked points at
distance at most $\ell$, related to the so-called two-point function
\cite{AW95}. Equation (\ref{eq:wlrecur}) is nothing but a refinement
of the third relation of (\ref{eq:tetra}) (and we have
$E_0^{(4)}=R_1/t$). Remarkably, it has an explicit solution
\cite[section 4.1]{geod}. It furthermore looks surprisingly analogous
to (yet different from) the ``first string equation''
(\CicutaMolinari.2.14). This still mysterious analogy is much more
general, as one may refine equations (\ref{eq:rsthm}) into discrete
recurrence equations involving the distance, similar to the string
equations for the one-matrix model, and having again explicit
solutions \cite{geod,PDF05}.

The correspondence between maps and trees has sparked an active field
of research between physics, combinatorics and probability theory,
devoted to the study of the geometry of large random maps, see for
instance \cite{Mier09} and references therein.


\end{document}

%% file: feynman2.pstex_t
\begin{picture}(0,0)%
\includegraphics{feynman2.pstex}%
\end{picture}%
\setlength{\unitlength}{3947sp}%
\begingroup\makeatletter\ifx\SetFigFont\undefined%
\gdef\SetFigFont#1#2#3#4#5{%
  \reset@font\fontsize{#1}{#2pt}%
  \fontfamily{#3}\fontseries{#4}\fontshape{#5}%
  \selectfont}%
\fi\endgroup%
\begin{picture}(11730,3431)(1636,-4660)
\put(2401,-2911){\makebox(0,0)[b]{\smash{{\SetFigFont{12}{14.4}{\familydefault}{\mddefault}{\updefault}{\color[rgb]{0,0,0}$i$}%
}}}}
\put(2401,-2161){\makebox(0,0)[b]{\smash{{\SetFigFont{12}{14.4}{\familydefault}{\mddefault}{\updefault}{\color[rgb]{0,0,0}$j$}%
}}}}
\put(2401,-3961){\makebox(0,0)[b]{\smash{{\SetFigFont{20}{24.0}{\familydefault}{\mddefault}{\updefault}{\color[rgb]{0,0,0}$M_{ij}$}%
}}}}
\put(2401,-4561){\makebox(0,0)[b]{\smash{{\SetFigFont{20}{24.0}{\familydefault}{\mddefault}{\updefault}{\color[rgb]{0,0,0}(a)}%
}}}}
\put(7201,-3961){\makebox(0,0)[b]{\smash{{\SetFigFont{20}{24.0}{\familydefault}{\mddefault}{\updefault}{\color[rgb]{0,0,0}$M_{ij} M_{jk} M_{ki}$}%
}}}}
\put(7201,-4561){\makebox(0,0)[b]{\smash{{\SetFigFont{20}{24.0}{\familydefault}{\mddefault}{\updefault}{\color[rgb]{0,0,0}(b)}%
}}}}
\put(7351,-2161){\makebox(0,0)[lb]{\smash{{\SetFigFont{12}{14.4}{\familydefault}{\mddefault}{\updefault}{\color[rgb]{0,0,0}$k$}%
}}}}
\put(7351,-2911){\makebox(0,0)[lb]{\smash{{\SetFigFont{12}{14.4}{\familydefault}{\mddefault}{\updefault}{\color[rgb]{0,0,0}$j$}%
}}}}
\put(6676,-2536){\makebox(0,0)[rb]{\smash{{\SetFigFont{12}{14.4}{\familydefault}{\mddefault}{\updefault}{\color[rgb]{0,0,0}$i$}%
}}}}
\put(12001,-2161){\makebox(0,0)[b]{\smash{{\SetFigFont{12}{14.4}{\familydefault}{\mddefault}{\updefault}{\color[rgb]{0,0,0}$j=k$}%
}}}}
\put(12001,-2911){\makebox(0,0)[b]{\smash{{\SetFigFont{12}{14.4}{\familydefault}{\mddefault}{\updefault}{\color[rgb]{0,0,0}$i=l$}%
}}}}
\put(12001,-3961){\makebox(0,0)[b]{\smash{{\SetFigFont{20}{24.0}{\familydefault}{\mddefault}{\updefault}{\color[rgb]{0,0,0}$\langle M_{ij} M_{kl} \rangle$}%
}}}}
\put(12001,-4561){\makebox(0,0)[b]{\smash{{\SetFigFont{20}{24.0}{\familydefault}{\mddefault}{\updefault}{\color[rgb]{0,0,0}(c)}%
}}}}
\end{picture}%